\documentclass[twocolumn,APS]{revtex4}

\begin{document}
 \title{The ground-state of General Relativity, Topological Theories and Dark Matter}
 \author{M\'{a}ximo Ba\~{n}ados  \\  Departamento de F\'{\i}sica,\\
 P. Universidad Cat\'{o}lica de Chile, Casilla 306, Santiago 22,Chile. \\
 {\tt maxbanados@fis.puc.cl}}

\begin{abstract}
We suggest a limit of Einstein equations incorporating the state $g_{\mu\nu}=0$ as a solution. The large scale behavior of this theory has interesting properties. For a spherical source, the velocity profile for circular motions is of the form observed in galaxies (approximately flat). For FRW cosmologies, the Friedman equation contains an additional contribution in the matter sector.
\end{abstract}

\maketitle

It has long been conjectured that the state $g_{\mu\nu}(x)=0$ should play a role in gravitation \cite{Witten88,Horowitz,Giddings}.  Since in Einstein gravity energies, and not just energy differences, are the sources of curvature it is particularly relevant to determine the true ground state.

As discussed in \cite{Horowitz}, general relativity can be studied at $g_{\mu\nu}=0$ by considering the metric and connection as independent variables. We shall argue here that the Einstein tensor evaluated at zero metric may be non-zero.  Thus, Einstein equations having $g_{\mu\nu}=0=T_{\mu\nu}$ as a solution read,
\begin{equation}\label{new0}
G_{\mu\nu}(g) = G_{\mu\nu}(0) + 8\pi G \, T_{\mu\nu}.
\end{equation}
$G_{\mu\nu}(0)$ is not a fixed object and does not conflict with Bianchi identities (see below). Unexpectedly, the new piece $G_{\mu\nu}(0)$ induces changes in the gravitational potential similar to dark matter. For recent discussions on dark matter see \cite{Spergel,Bradac}. Numerical studies for the density profiles have been done in \cite{NFW}. Supersymmetric particles are the best candidates for dark matter (for a review see \cite{Jungman}).  The lack of direct observation has also led to alternative descriptions \cite{Sanders,Bekenstein,Ferreira,Carroll}. Discussions on the role of metric invertibility have appeared in \cite{Wilczek,Deser}.

It is important to mention that formulations of general relativity incorporating  $g_{\mu\nu}=0$ and without adding extra pieces do exist. The simplest \cite{Horowitz} uses the vielbein $e^a$ and a first order ($3-$form) Einstein tensor $G_a=\epsilon_{abcd}R^{ab}\wedge e^c$ which is zero at $e^a=0$. $G_a$ and $G_{\mu\nu}$ carry the same information for invertible metrics but they clearly differ for non-invertible ones. An advantage of the vielbein formulation is that an action exists, but it is not quadratic around $e^a=0$.

Our first goal is to explore the values of the Einstein tensor $G_{\mu\nu}$ at $g_{\mu\nu}(x)=0$.
We start by reviewing the analysis of \cite{Horowitz}. Recall
\begin{equation}\label{G}
G_{\mu\nu} = R_{\mu\nu}(\Gamma) - {1 \over 2} g_{\mu\nu}g^{\alpha\beta} R_{\alpha\beta}(\Gamma),
\end{equation}
where we have emphasized that $R_{\mu\nu}(\Gamma)$ depends only on the connection. If $g_{\mu\nu}$ is invertible, the metricity condition $g_{\mu\nu;\rho}=0$ yields the Christoffel symbol $\Gamma(g)$. On the contrary, if $g_{\mu\nu}(x)=0$ the metricity condition yields $0=0$. This means that the connection decouples from the metric and becomes an independent field. The ground state $g_{\mu\nu}=0$ is thus parameterized by a connection field denoted by $\Gamma_0(x)$.  In particular, at $g_{\mu\nu}=0$, the Ricci curvature $R_{\mu\nu}(\Gamma_0)$ is bounded and a well-defined function of $\Gamma_0(x)$ \cite{Horowitz}.

The second term in the Einstein tensor, ${1 \over 2}g_{\mu\nu}g^{\alpha\beta} R_{\alpha\beta}(\Gamma)$, is more delicate. The curvature factor is well-defined but the pre-factor becomes ${0 \over 0 }$.  The value of this object at $g_{\mu\nu}(x)=0$ can only be defined by a limiting process $g_{\mu\nu}(x)\rightarrow 0$.

We are led to consider the creation of a Riemannian metric $g_{\mu\nu}(x)$ starting from $g_{\mu\nu}(x)=0$.  Once the metric is created, it must satisfy Einstein equations. But the process of creating it is probably much more complicated. To put an analogy, consider the creation of a particle from two photons.  The dynamics of the particle, after being created, may be controlled by Newton equations. But the creation process itself is described by quantum field theory. Nevertheless, classical dynamics does provide interesting information. From the relativistic energy $E(p) = \sqrt{m^2 + p^2}\approx m + {p^2 \over 2m} + \cdots$ one discovers that $E(0)=m$ is the energy to create the particle.  The goal of this paper is to push a similar analysis for the Einstein tensor.  The tensor $G_{\mu\nu}(0)$ entering in (\ref{new0}) is an energy-momentum  associated to the transition from $g_{\mu\nu}=0$ to $g_{\mu\nu}\neq 0$. We shall not attempt to find a theory describing this process. This is well beyond the scope of this paper. But we shall argue that in some cases $G_{\mu\nu}(0)$ can be computed with interesting consequences.

Consider paths in the space of metrics starting at $g_{\mu\nu}=0$. For every given path, the value of $\lim_{g\rightarrow 0}\left[ {1 \over 2}g_{\mu\nu}g^{\alpha\beta} R_{\alpha\beta}(\Gamma)\right]$ can be computed. Since this object has the structure ${0 \over 0}$ the result may be zero, finite or divergent. We work under the assumption that a process creating metrics from $g_{\mu\nu}=0$ exists.  Then we are led to assume that the relevant paths (with physical meaning) are such that  $\lim_{g\rightarrow 0}\left[ {1 \over 2}g_{\mu\nu}g^{\alpha\beta} R_{\alpha\beta}(\Gamma)\right]$ is zero or finite. This assumption is enough to guess its form.  This object is a rank-two symmetric tensor with dimensions of curvature. At $g_{\mu\nu}=0$ the only field is $\Gamma_0$ and there are not too many tensors with these properties. We postulate \footnote{Other symmetric objects can be built using the covariant derivative and the inverse of $R_{\mu\nu}$. Those will be of higher order in derivatives and involve dimension full constants. We neglect them.} $\lim_{g\rightarrow 0} \left[ {1 \over 2 } g_{\mu\nu}g^{\alpha\beta} R_{\alpha\beta} \right] = \lambda' \, R_{\mu\nu}(\Gamma_0)$ where $\lambda'$ is a dimensionless parameter depending on the path.

Our proposal for the Einstein tensor at zero metric is then summarized as
\begin{equation}\label{G0}
G_{\mu\nu}(0)  = \lambda\, R_{\mu\nu}(\Gamma_0),
\end{equation}
where $\lambda = 1 - \lambda'$.

Let us now argue that the connection $\Gamma_0(x)$ is also a function of the path. The point is that the Christoffel symbol $\Gamma(g) \sim {1 \over 2}g^{-1}\partial g$ becomes also ${0 \over 0}$ at $g_{\mu\nu}(x)=0$ \footnote{$g_{\mu\nu}(x)=0$ at {\it all} points in spacetime also implies $\partial_\rho g_{\mu\nu}\rightarrow 0$.}, in consistency with the fact that the metricity condition becomes $0=0$. If at $g_{\mu\nu}=0$ the connection is $\Gamma_0(x)$ then continuity implies
\begin{equation}\label{path}
\Gamma_{0}(x) = \lim_{g\rightarrow 0}\Gamma(g),
\end{equation}
where the limit is taken along the chosen path.  In summary, $G_{\mu\nu}(0)$ is a function of the path describing the creation of the metric.

We now turn to the examples where symmetry principles together with Bianchi identities allow the determination of $G_{\mu\nu}(0)$.

\noindent {\bf Spherical sources and galactic rotation curves.} Consider, for example, the ``creation" of a spherically symmetric spacetime. The initial state is characterized by $g_{\mu\nu}(x)=0$ and a connection $\Gamma_0(x)$.  The final state is a spherically symmetric metric $g_{\mu\nu}(x)$, with its associated Christoffel connection $\Gamma(g)$.

If the final state is spherically symmetric it is reasonable to assume that   $\Gamma_0(x)$ must also be spherically symmetric. This can be understood as a selection rule between the final and initial state. The general form for a spherically symmetric metric is,
\begin{equation}\label{Sch}
ds^2 = - A^2(r) dt^2 + B^2(r) dr^2 + C^2(r) d\Omega^2,
\end{equation}
and the non-zero curvature components are,
\begin{eqnarray}
{1 \over c^2}\,  R_{tt}(g) &=& {A \over B}\left({A' \over B}\right)' + {2A \over C}{A'C' \over B^2} \label{R1}\\
  R_{rr}(g) &=& - {A'' \over A} + {A'\over A}{B' \over B} - {2C'' \over C} + {2B' \over B }{C' \over C}  \label{R2}\\
  R_{\theta\theta}(g) &=& -{C \over A}{C'A' \over B^2} - {C \over B}\left({C' \over B}\right)' - {C'^2 \over B^2} + 1 \label{R3}
\end{eqnarray}
plus $R_{\phi\phi}=\sin(\theta)^2 R_{\theta\theta}$. We shall evaluate $R_{\mu\nu}(\Gamma_0)$ entering in (\ref{G0})  by taking the limit $A,B,C\rightarrow 0$ of these expressions. Note that the gauge $C=r$ cannot be fixed too soon because one looses the possibility of exploring $g_{\mu\nu}\rightarrow 0$. Interestingly, this residual gauge freedom will play an important role below.

As expected, $R_{\mu\nu}(\Gamma_0)$ has the structure ${0 \over 0}$ as $A,B,C\rightarrow 0$.  Recall that $A(r)\rightarrow 0$ for all $r$ implies $A'(r) \rightarrow 0$ too (assuming functions $C^{\infty}$).  However, $A'$ is not a scalar under the residual radial redefinitions of (\ref{Sch}).  In principle, one could do radial re-scalings and keep $A'(r)$ finite.  An invariant way to represent $A' \rightarrow 0$ avoiding this ambiguity is
\begin{equation}\label{limA}
 {A' \over B}\rightarrow 0.
\end{equation}
This is equivalent to ${dA(\rho)\over d\rho} \rightarrow 0$ with $\rho$ the proper radial coordinate.  We shall postulate that during the transition (\ref{limA}) holds.    The same analysis is valid for $C(r)$ and thus $C'/B \rightarrow 0$.  Coordinate invariance of the limit then fixes some of the quotients appearing in (\ref{R2}) to be zero.

Consider now quotients of the form ${A' \over A}$,${C' \over C}$ appearing in $R_{rr}(\Gamma_0)$. These have a completely different interpretation.  On the one hand $R_{rr}$ is not a scalar under radial rescalings.  On the other $\lim_{A\rightarrow 0} A'/A$ cannot be fixed to any particular value.  As a toy example consider $A=a + br$ with $\lim_{A\rightarrow o} A'/A = \lim_{a,b\rightarrow 0}\, b/(a+br)$. If we set $b\rightarrow 0$ first and then $a \rightarrow 0$ we obtain zero. In the reverse order we obtain $1/r$.  Setting $a=b \rightarrow 0$ yields  $1/(1+r).$ In summary, the limit $\lim {A' \over A}$ is finite but completely arbitrary depending on the path near $A=0$. For a general function $A(r) = a_0 + a_1r + a_2 r^2 + \cdots$ the limit $A'/A$ is fully arbitrary.

In summary, for spherical symmetry $R_{\mu\nu}(\Gamma_0)$ has  the form,
\begin{equation}\label{R01}
R_{tt}(\Gamma_0)=0, \ \ \ \ \  R_{rr}(\Gamma_0)=q(r), \ \ \ \ \
R_{\theta\theta}(\Gamma_0)=1,
\end{equation}
(plus $R_{\phi\phi}=\sin^2(\theta)$) where $q(r)$ is an arbitrary function built from $A'/A, C'/C$, etc. As we now show, Einstein equations fix $q(r)$.

We insert (\ref{R01}) in (\ref{new0}) and solve for the functions $A,B,q$ in the radial gauge $C=r$.  Consider first the `exterior' problem with $T_{\mu\nu}=0$. Since $R_{tt}(\Gamma_0)=0$, the $tt$ Einstein equation can easily be solved obtaining the usual Schwarzschild formula for $B^2(r)$
\begin{equation}\label{B}
B^2(r) = \left(  1 - {2m_{0}G \over c^2r} \right)^{-1}, \nonumber
\end{equation}
where $m_{0}$ is an integration constant associated to the total `visible' matter (see below). The $rr$ and $\theta\theta$ Einstein equations are more complicated. It is convenient to redefined the metric function $A^2(r)$ as
\begin{equation}\label{A}
A^2(r) =\left(1- {2m_{0}G \over c^2r}\right) \, w^2(r).
\end{equation}
The $rr$ and $\theta\theta$ Einstein equations (\ref{new0}) become,
\begin{eqnarray}
  {2w' \over w}   &=& \lambda\, r\, q  \label{sp1} \\
 r\left(r-{2m_{0}G \over c^2} \right)w'' + \left(r + {m_{0} G\over c^2} \right) w' &=& \lambda\, w. \label{sp2}
\end{eqnarray}
These are two equations for two unknowns $w(r)$ and $q(r)$. As promised, all unknown functions are thus determined by the equations of motion.

The exact solution to (\ref{sp2}) is available but we restrict the discussion here to the non-relativistic limit $c\rightarrow \infty$ and expand the solution in powers of $1/c^2$.  The geometry outside a Galaxy is approximately flat thus $w(r)$ must be of the form $w(r) = 1 + $corrections.  This implies that $\lambda$ must be small and scale with $c$ as
\begin{equation}\label{k}
\lambda = {k \over c^2}.
\end{equation}
The solution for $w(r)$ becomes
\begin{equation}\label{w}
w(r) = 1 + {1 \over 2c^2} \left( C_0 \ln(r/r_0) + k \ln^2(r/r_0) \right) + \cdots
\end{equation}
where $C_0$ and $r_0$ are integration constants.  The term proportional to $C_0$ is the log branch which give rise to Galactic flat rotation curves. In fact, a test particle in circular motion at distance $r$ from the center has a speed,
\begin{equation}\label{v}
v^2(r) = r \Phi',
\end{equation}
where the Newtonian potential $\Phi$ is determined from $A^2 \simeq 1 + 2\Phi(r)/c^2 + \cdots$.  Inserting (\ref{A}) and (\ref{w}) the velocity (\ref{v}) is
\begin{equation}\label{ext}
v^2(r) =  {m_{0}G \over r} + {C_0 \over 2} + k\, \ln(r/r_0).
\end{equation}
The first term is the usual `visible' matter contribution.  This is now superposed with a constant contribution $C_0$ characteristic of flat rotation curves. The third term is a correction proportional to $k$.  We see that the dark halo is independent of the visible matter, at least to this order. Adjusting $C_0,r_0$ and $k$, (\ref{ext}) has the right shape for realistic situations.

Let us now briefly consider the interior solution with $T_{\mu\nu}\neq 0$ parameterized by a mass density $\mu(r)$ and pressure $p(r)$. This is relevant for motions inside a galaxy. As usual for interior solutions, it is convenient to parameterize the metric functions as,
\begin{eqnarray}
  B^2 &=& \left( 1 - {2m(r)G \over c^2 r} \right)^{-1},  \\
  A^2 &=& e^{2\Phi(r)/c^2} \simeq 1 + {2\Phi(r) \over c^2} + \cdots, \label{AA}
\end{eqnarray}
where $m(r)$ and $\Phi(r)$ are undetermined functions. In the non-relativistic limit, the $tt$ Einstein equation (\ref{new0}) and energy-momentum conservation give $
m' = 4\pi \mu r^2$ and $p' = -\mu \Phi'$ respectively,  coinciding with the usual formulae. Next, the $rr$ Einstein equation fixes the value of $q(r)$, which we don't display here.  Finally, using (\ref{k}), the $\theta\theta$ equation becomes
\begin{equation}\label{matter}
(r\Phi')' - \left( {m(r)G \over r} \right)' - {k \over r} = 0,
\end{equation}
from where $\Phi(r)$ can be determined. The velocity profile becomes,
\begin{equation}\label{int}
v^2(r) = {m(r)G \over r} + {C_0 \over 2} +  k\, \ln(r/r_0).
\end{equation}
where we have chosen an integration constant to coincide with (\ref{ext}). Continuity between the exterior, (\ref{ext}), and interior, (\ref{int}), solutions is thus manifest adjusting $m_0$ to be the total visible matter.

~

\noindent {\bf Friedman-RW models.} Our next example is a cosmological model described by the metric
\begin{equation}\label{cos}
d^2 = -N^2(t) dt^2 + a(t)^2 \left( {dr^2 \over 1-kr^2} + r^2 d\Omega^2 \right)
\end{equation}
We would like to find the evolution equation for $a(t)$ implied by (\ref{new0}). The steps to solve this problem are exactly the same as those for the spherical source. The non-zero independent components for the Ricci tensor are,
\begin{eqnarray}
R_{tt}(g)  &=& -{3\ddot a \over a} + {3\dot N \over N} {\dot a \over a} \\
R_{rr}(g) &=& {1 \over 1-kr^2} \left[- {a\dot a \dot N \over N^3} + { a \ddot a  \over N^2} + {2\dot a^2 \over N^2} + 2k \right].
\end{eqnarray}
and we study their limit as $N,a \rightarrow 0$. Imposing  the proper condition $\lim {\dot a \over N}=0$ and leaving ${\dot a \over a}$ and ${\dot N \over N}$ as arbitrary functions, $R_{\mu\nu}(\Gamma_0)$ is
\begin{equation}\label{cosR}
R_{tt}(\Gamma_{0}) = q(t), \ \ \ \ \ \ \  R_{rr}(\Gamma_{0}) = {2k \over 1-kr^2}
\end{equation}
where $q(t)$ is an arbitrary function of time. Replacing (\ref{cosR}) in (\ref{new0}), in the gauge $N(t)=1$, the independent Einstein equations (\ref{new0}) including an ideal fluid characterized by a pressure $p(t)$ and density $\rho(t)$ are
\begin{eqnarray}
{\dot a^2  \over a^2} + {k \over a^2}  &=& {1 \over 3}\lambda\, q(t) + {8\pi G \over 3} \rho(t), \label{c1} \\
2a \ddot a + \dot a^2 + k &=& -2 \lambda\,k  - 8\pi G a^2 p(t),  \label{c2}\\
(\rho a^3)^. &=& -p (a^3)^. \label{c3}
\end{eqnarray}
If $\lambda$ was zero, (\ref{c1}) is the Friedman equation, and (\ref{c2}) is related to it by a Bianchi identity. But, if $\lambda \neq 0$, (\ref{c1}) fixes $q(t)$ and (\ref{c2}) is independent becoming the dynamical equation for $a(t)$. Again, all arbitrary functions have been fixed.

Multiplying (\ref{c2}) by $\dot a$ it can be integrated once,
\begin{equation}\label{FRW}
{\dot a^2 \over a^2} + {k(1+2\lambda) \over a^2} = {8\pi G \over 3} \rho + {C \over a^3},
\end{equation}
where $C$ is an arbitrary integration constant.

This equation differs from the usual Friedman equations in two respects.  First, the  curvature term $1/a^2$ appears ``renormalized".  Second, there is an extra contribution $C/a^3$ which increases the visible density $\rho$.  The exponent $a^{-3}$ yields the correct  equation of state for dark matter.  Thus, adjusting the integration constant $C$, this equation becomes consistent with observations without adding extra components.

To conclude,  we have explored dark matter as a topological contribution to Einstein equations.  Since dark matter is known not to interact with ordinary matter this is an  attractive possibility.  Both galactic flat rotation curves and extra matter in the Friedman equation arise in natural way within this framework. It is interesting to observe that $G_{\mu\nu}(0)$ does not explicitly introduce extra matter. The mechanism is different. The first order Einstein equations fix $\Gamma_0(x)$. The second order Einstein equations, normally related to the first order ones by Bianchi identities, now become independent having extra integration constants.

There are several open problems. Most importantly, the general dynamical equation for  $\Gamma_0$ which is required for general problems like fluctuations. We shall come back to this problem elsewhere.

~

I would like to thank T. Andrade, E. Ayon-Beato, A.Clocchiatti, S. Deser, G.Galaz, M. Garc\'{\i}a del Moral, A.Gomberoff, M. Henneaux, L. Houart, T. Jacobson, C.Martinez, S. Theisen, E. Vagenas, J.Zanelli for useful comments and A. Reisenegger for many conversations and remarks which contributed enormously to this work. Useful comments on an earlier version of the manuscript by S. Carlip and S. Theisen are also gratefully acknowledge.  The author was partially supported by Fondecyt Grants (Chile) \#1060648 and \#1051084.


\begin{thebibliography}{10}

\bibitem{Witten88}
  E.~Witten,
  Nucl.\ Phys.\ B {\bf 311}, 46 (1988).

\bibitem{Horowitz}
  G.~T.~Horowitz,
  Class.\ Quant.\ Grav.\  {\bf 8}, 587 (1991).

\bibitem{Giddings}
  S.~B.~Giddings,
  Phys.\ Lett.\ B {\bf 268}, 17 (1991).

\bibitem{Wilczek}
  F.~Wilczek,
  Phys.\ Rev.\ Lett.\  {\bf 80}, 4851 (1998)

\bibitem{Deser}
  S.~Deser,
  Class.\ Quant.\ Grav.\  {\bf 23}, 4539 (2006)

\bibitem{Jungman}
  G.~Jungman, M.~Kamionkowski and K.~Griest,
  Phys.\ Rept.\  {\bf 267}, 195 (1996)
  [arXiv:hep-ph/9506380].

\bibitem{Sanders}
  R.~H.~Sanders,
  arXiv:astro-ph/0601431.

\bibitem{Bekenstein}
  J.~D.~Bekenstein,
  Phys.\ Rev.\ D {\bf 70}, 083509 (2004)

\bibitem{Ferreira}
  C.~Skordis, D.~F.~Mota, P.~G.~Ferreira and C.~Boehm,
  Phys.\ Rev.\ Lett.\  {\bf 96}, 011301 (2006)

\bibitem{Spergel}
  D.~N.~Spergel {\it et al.},
  arXiv:astro-ph/0603449.

\bibitem{Bradac}
  M.~Bradac {\it et al.},
  arXiv:astro-ph/0608408.

\bibitem{NFW}
  J.~F.~Navarro, C.~S.~Frenk and S.~D.~M.~White,
  Astrophys.\ J.\  {\bf 462}, 563 (1996)
  [arXiv:astro-ph/9508025].


\bibitem{Carroll}
  S.~M.~Carroll et al,   
  Phys.\ Rev.\ D {\bf 71}, 063513 (2005)

\end{thebibliography}
 \end{document}